\documentclass[smallextended]{svjour3}       
\smartqed  
\usepackage{graphicx}
\usepackage{amssymb}
\usepackage{amsmath}
\usepackage{amsfonts}
\usepackage{mathrsfs}
\usepackage{bm}
\usepackage{color}
\usepackage{mathbbol}
\usepackage{appendix}
\usepackage{enumerate}
\usepackage{multirow}
\begin{document}

\title{Quantum cloning attacks against PUF-based quantum authentication systems
}


\author{Yao Yao       \and
       Ming Gao       \and
       Mo Li          \and
       Jian Zhang
}


\institute{Yao Yao \and Ming Gao \and Mo Li \and Jian Zhang \at
              Microsystems and Terahertz Research Center, China Academy of Engineering Physics, Chengdu Sichuan 610200, China\\
              Institute of Electronic Engineering, China Academy of Engineering Physics, Mianyang Sichuan 621999, China\\
              \email{yaoyao@mtrc.ac.cn, limo@mtrc.ac.cn}           
}

\date{Received: date / Accepted: date}

\maketitle

\begin{abstract}
With the advent of Physical Unclonable Functions (PUFs), PUF-based quantum authentication systems (QAS) have
been proposed for security purposes and recently proof-of-principle experiment has been demonstrated.
As a further step towards completing the security analysis, we investigate quantum cloning attacks against
PUF-based quantum authentication systems and prove that quantum cloning attacks outperform the so-called
challenge-estimation attacks. We present the analytical expression of the false accept probability by use of
the corresponding optimal quantum cloning machines and extend the previous results in the literature.
In light of these findings, an explicit comparison is made between PUF-based quantum authentication systems and quantum
key distribution (QKD) protocols in the context of cloning attacks. Moreover, from an experimental perspective,
a trade-off between the average photon number and the detection efficiency is discussed in detail.
\end{abstract}

\section{Introduction}\label{sec1}
Modern cryptographic applications, such as identification and authentication, rely on \textit{mathematical} one-way
functions, which provide significant asymmetry between a black-box inverter and a non-black-box extractor.
However, they are confronted with serious challenges from both practical and fundamental aspects. On one hand,
the enormous development has recently been made in the field of network parallel computation and rapid reverse-engineering technology;
on the other hand, a complete proof of security is still missing although the cryptographic primitives employed nowadays are believed to be secure.

The concept of physical unclonable function (PUF) was introduced by R. Pappu et al. to address all these security issues \cite{Pappu2001,Pappu2002}.
In contrast to mathematical one-way functions, a PUF is a physical entity that is embodied in a specific physical structure and is easy to evaluate
but hard to characterize \cite{Gassend2003}. As its name suggests, an individual PUF must be easy to be fabricated but practically infeasible to be duplicated,
even given the exact manufacturing process that produced it. In this sense, it is the physical analog of a mathematical one-way function.
Since the advent of the definition of PUF, a great deal of PUF or PUF-like proposals and implementations have been raised as cryptographic primitives for
security purposes, such as Optical PUFs \cite{Pappu2001,Pappu2002}, Coating PUFs \cite{Tuyls2006}, Arbiter PUFs \cite{Lim2004},
Ring Oscillator PUFs \cite{Suh2007}, SRAM PUFs \cite{Guajardo2007}, etc. For more details, we refer the readers to Ref. \cite{Maes2012}.

Due to the desirable properties of PUF, such as unclonablity and unpredictability, it naturally leads us to design authentication or
anti-counterfeiting protocols based on PUF devices. For instance, in the seminar paper by R. Pappu et al. \cite{Pappu2001,Pappu2002},
one of the most significant demonstrations is to design authentication and anti-countering protocol based on PUF made of a three-dimensional random scattering medium.
It is worth noting that the security of majority of PUF implementations is guaranteed by the complex structure of the corresponding physical objects, e.g.,
the random distribution of scatterers in optical medium \cite{Pappu2001,Pappu2002}.
Therefore, these \textit{classical} PUF-based authentication protocols are theoretically breakable if infinite computing power
is assumed. To achieve a higher level of security, B. \u{S}kori\'{c} proposed a new type of PUF, the quantum-readout PUF, which
combines the classical optical PUF with the quantum challenges \cite{Skoric2012}. Recently the first proof-of-principle experiment
of PUF-based quantum-secure authentication has been demonstrated by S. A. Goorden et al. \cite{Goorden2014},
attracting a great deal of attention from researchers \cite{Miller2015}.

However, the security analysis of PUF-based quantum-secure authentication protocols is still not more than mature or integrate due to
its relative youth. To our best knowledge, only \textit{classical} attacks, such as the challenge estimation attacks, have been considered
\cite{Skoric2013a,Skoric2013b}. As a further step towards completing the security analysis, in this work we proposed the quantum cloning attacks,
the first \textit{quantum} attack scenario against quantum-secure authentication systems and investigate its
effects upon the false-accept probability. The outline of the reminder of this paper is as follows: In Sect. \ref{sec2}, we provide
a brief review of technical preliminaries of quantum-readout PUFs and notations used throughout the paper.
In the following Sect. \ref{sec3}, we investigate quantum cloning attacks against PUF-based quantum authentication systems
and prove that quantum cloning attacks outperform the challenge-estimation attacks. In Sect. \ref{sec4},
we extend our consideration to other quantum cloning attacks in the framework of quantum cryptography.
Finally, Sect. \ref{sec5} is devoted to the discussion and conclusion.

\section{The model of PUF-based quantum authentication systems}\label{sec2}
\subsection{The challenge and response behavior of optical PUFs}
\begin{figure}[htbp]
\begin{center}
\includegraphics[width=.80\textwidth]{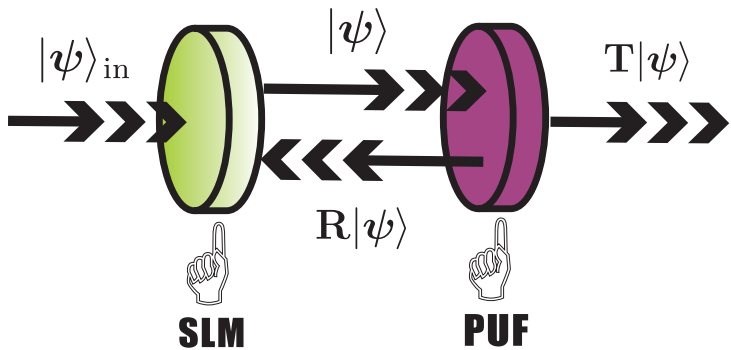} {}
\end{center}
\caption{The sketch of PUF-based quantum authentication system. The PUF device can be
characterized as the scattering matrix $S$, which relates to the transmission and reflection matrices $T$ and $R$
respectively, according to Eq. (\ref{S-matrix}). The spatial light modulator (SLM) is used for
creating the desired challenge states by modulating the phase of the initial laser beam.}
\label{fig1}
\end{figure}

In this subsection, we mainly focus on the modeling of optical PUFs \cite{Pappu2001,Pappu2002}. Note that the transport of light
through a multiple-scattering medium can be described by means of the scattering matrix or $S$-matrix \cite{Beenakker1997}.
Therefore, the key point is to view the PUF as a \textit{black-box}. If the collisions between photons and
scatterers are further assumed to be completely elastic, the behavior of such a black-box can be fully characterized by
a unitary transformation, which is represented by a unitary $S$-matrix (see Fig. \ref{fig1}).

To describe the functionality of quantum-readout PUFs \cite{Skoric2012,Goorden2014}, three types of quantum states are involved: the challenge state,
reflected state and transmitted state \cite{Skoric2012,Goorden2014}. All these state we are dealing with have
two intertwined degree of freedoms: the \textit{internal} degree of freedom characterizes the optical properties
(e.g., spin or polarization) but the \textit{external} degree of freedom depicts the directivity of light (e.g., direction of motion),
which is mathematically denoted as a tensor product of Hilbert spaces $\mathcal{H}=\mathcal{H}_{\text{ext}}\otimes\mathcal{H}_{\text{int}}$.
Thus, any quantum state $|\psi\rangle\in\mathcal{H}$ can be represented as
\begin{align}
|\psi\rangle=|\text{outgoing}\rangle\otimes|\psi_1\rangle+|\text{incoming}\rangle\otimes|\psi_2\rangle
=\left(\begin{matrix}
|\psi_1\rangle \\
|\psi_2\rangle \\
\end{matrix}\right),
\end{align}
where $|\text{incoming}\rangle$, $|\text{outgoing}\rangle\in\mathcal{H}_{\text{ext}}$, $|\psi_1\rangle$, $|\psi_2\rangle\in\mathcal{H}_{\text{int}}$
and here \textit{incoming} or \textit{outgoing} is with respect to the sender.

Equipped with these notations, the interaction between the quantum state and the optical PUF can be expressed as (see Fig. \ref{fig1})
\begin{align}
\left(\begin{matrix}
|\psi'_1\rangle \\
|\psi'_2\rangle \\
\end{matrix}\right)
=S\left(\begin{matrix}
|\psi_1\rangle \\
|\psi_2\rangle \\
\end{matrix}\right)
=\left(\begin{matrix}
T & -R^{\dagger} \\
R & T^{\dagger}
\end{matrix}\right)
\left(\begin{matrix}
|\psi_1\rangle \\
|\psi_2\rangle \\
\end{matrix}\right),
\label{S-matrix}
\end{align}
where the operators $R$ and $T$ denote the reflection and transmission submatrices respectively. Due to the unitarity
of the $S$-matrix, we have the relation $RR^{\dagger}+TT^{\dagger}=\mathbb{1}$.

\subsection{The authentication protocol}\label{sec2.2}
We only consider the authentication protocol based on the reflected states by assuming $T=0$, following the experimental implementation in \cite{Goorden2014}.
Note that in this case there is no loss of generality since any transmitted state can always be re-routed to become part of the reflected state \cite{Skoric2012}.
The detailed protocol consists of two steps:

$\bullet$ \textbf{Enrollment phase}: Alice assigns an identity code $I$ to a specific quantum-readout PUF and the challenge-response
state pairs are measured with as much light as needed. This procedure also yields the reflection matrix $R$ \cite{Skoric2012}.
The challenge along with the corresponding response is stored in a challenge-response database. Then this PUF is delivered to Bob.

$\bullet$ \textbf{Verification phase}: At some time Bob claims to have access to the quantum-readout PUF with the identifier $I$.
Alice looks up the identity code $I$ in the database and finds the corresponding matrix $R$. Then she initializes two counters
$n_1$ and $n_2$ to zero and repeats the following procedure $m$ times:

\begin{enumerate}
\renewcommand{\labelenumii}{(\arabic{enumii})}
    \item Alice randomly prepares a state $|\psi\rangle$ with uniform distribution and sends it to Bob as a challenge state.
    \item Within a permitted amount of time, Alice will obtain two distinctive responses: nothing or a reflected state $|\chi\rangle$.
If she receives a state, then
    \begin{enumerate}
        \item The first counter $n_1$ is increased by one;
        \item Since the reflection matrix $R$ is known, the valid reflected state should be $|\chi_{\psi}\rangle=R|\psi\rangle$.
Alice performs the measurement defined by the projection operators $\{E_1=|\chi_{\psi}\rangle\langle\chi_{\psi}|,E_0=\mathbb{1}-E_1\}$
onto the received state $|\chi\rangle$, where the outcome `1' and `0' corresponds to $E_1$ and $E_0$ respectively. If the outcome is
`1', the second counter $n_2$ is increased by one.
    \end{enumerate}
\end{enumerate}

\noindent Finally, if the fraction $n_1/m$ of the responses is not consistent with the expected noise level, the authentication is aborted;
if $n_2\geq(1-\varepsilon)n_1$ then Alice is convinced that she has probed the PUF with the identifier $I$. Here $\varepsilon$ is a robustness
parameter denoting the tolerable fraction of wrong responses.

\section{Quantum cloning attacks}\label{sec3}
\subsection{Universal quantum cloning machine and false-accept probability}
In Ref. \cite{Skoric2012,Skoric2013a,Skoric2013b}, the challenge-estimation attacks or so-called intercept-resend attacks
have been considered for security analysis of PUF-based quantum authentication systems, which are regarded as the strongest
\textit{classical} attacks (the reason for this claim will be clear below). Here we propose the first quantum attack scenario
resorting to the optimal quantum cloning machines. Before presenting main results, two observations attract our attention:
(i) from the experimental perspective, a weak coherent laser source is usually employed instead of the fragile single-photon
states, which implies that multi-photon cases should be taken into account since every photon in these pulses will be modulated
by the spatial light modulator (SLM) with the same configuration (see Fig. \ref{fig1}); (ii) the challenge state is chosen at random, which indicates that
no specific state is superior to others and thus the universal quantum cloning machine is a more appropriate choice for our topic.

A $N\rightarrow M$ quantum cloning machine (QCM) is a completely positive map
$\mathcal{L}: \mathcal{B}(\mathcal{H}^{\otimes N})\rightarrow \mathcal{B}(\mathcal{H}^{\otimes M})$, which maps
input density operators of $N$ identical pure originals into output density operators of $M$ clones. In general, the average
fidelity of the cloning process is
\begin{align}
\mathcal{F}_{\text{QCM}}(N,M)=\sum_{k=1}^M\int_{\psi}\frac{1}{M}\left\langle\psi\left|\text{tr}_{\bar{k}}
\mathcal{L}\left(\psi^{\otimes N}\right)\right|\psi\right\rangle d\psi,
\end{align}
where $\mathcal{L}(\psi^{\otimes N})=\rho_{1,2,...,M}$ and $\text{tr}_{\bar{k}}$ denotes
the trace with respect to all the subsystems but $k$. Note that
for a $d$-dimensional Hilbert space $\mathcal{H}$, the set of pure states is isomorphic with the complex projective space $\mathbb{C}P^{d-1}$.
On this space there exists a unique natural measure $d\psi$, induced by the uniform Haar measure $d\mu(U)$ on the unitary group $U(d)$ \cite{Zyczkowski2001,Zyczkowski2008}.
If we restrict our consideration to the universal quantum cloning machine (UQCM),
then the $M$ clones are all in the same states since the output of the cloning machine is supported on the symmetric subspace \cite{Werner1998}.
Therefore, the reduced state of one copy can be defined using the notation $\mathcal{L}^s=\text{tr}_{\bar{1}}\mathcal{L}=\cdots=\text{tr}_{\bar{M}}\mathcal{L}$.
Now the average fidelity can be expressed as
\begin{align}
\mathcal{F}_{\text{UQCM}}(N,M)=\int_{\psi}\left\langle\psi\left|
\mathcal{L}^s\left(\psi^{\otimes N}\right)\right|\psi\right\rangle d\psi.
\end{align}

Equipped with these notations, we investigate the following attack model: the adversary first intercepts the incoming $N$-photon pulse $|\psi\rangle^{\otimes N}$ and
then applies the universal $N\rightarrow M$ \textit{qudit} cloner. Then in the language of the experiment \cite{Goorden2014}, the collective unitary transformations $R\otimes\cdots\otimes R=R^{\otimes M}$ is automatically performed by the optical PUF and the final state $R^{\otimes M}\mathcal{L}\left(\psi^{\otimes N}\right)R^{\dagger\otimes M}$ is sent back to Alice.
Note that here it is assumed that $R$ is a public information or the adversary has had access to the PUF temporarily in the past \cite{Skoric2012}.
We are now in a position to state the main result of this section (see Appendix \ref{appendix1} for more details).

\begin{theorem}
Under the quantum cloning attack by $N\rightarrow M$ UQCM, the adversary's success probability or the authentication system's false-accept probability
for a single copy is upper bounded by
\begin{align}
\mathcal{P}_{\text{accept}}\leq\frac{M-N+N(M+d)}{M(N+d)},
\end{align}
where $d=\text{dim}\mathcal{H}$, which is determined by the SLM configuration.
\label{T1}
\end{theorem}

\textit{Proof.} Following the protocol described in Sect. \ref{sec2.2}, the projections of each quanta (e.g., photon in \cite{Goorden2014})
are independent events in the detection process. Therefore, we can trace out all other subsystems when considering the single-copy case.
Mathematically, the adversary's success probability or the authentication system's false-accept probability
for a single copy is
\begin{align}
\mathcal{P}_{\text{accept}|\psi}&=\left\langle\chi_\psi\left|R\mathcal{L}^s(\psi^{\otimes N})R^{\dagger}\right|\chi_\psi\right\rangle
=\left\langle\psi\left|R^{\dagger}R\mathcal{L}^s(\psi^{\otimes N})R^{\dagger}R\right|\psi\right\rangle \nonumber\\
&=\left\langle\psi\left|\mathcal{L}^s(\psi^{\otimes N})\right|\psi\right\rangle,
\end{align}
where the relation $RR^{\dagger}=\mathbb{1}$ is applied here. Note that $\mathcal{P}_{\text{accept}|\psi}$ is conditioned on a specific challenge state.
To eliminate this dependence, we can average over the challenge state space (e.g., the pure state space)
\begin{align}
\mathcal{P}_{\text{accept}}=\mathbb{E}_\psi P_{\text{accept}|\psi}=\int_{\psi}\left\langle\psi\left|
\mathcal{L}^s\left(\psi^{\otimes N}\right)\right|\psi\right\rangle d\psi=\mathcal{F}_{\text{UQCM}}(N,M).
\end{align}
Unitizing the results in \cite{Werner1998,Keyl1999}, the false-accept probability is upper bounded by the \textit{optimal} single-copy fidelity of UQCM
\begin{align}
\mathcal{P}_{\text{accept}}\leq\mathcal{F}^{\text{opt}}_{\text{UQCM}}(N,M)=\frac{M-N+N(M+d)}{M(N+d)}.
\end{align}
This completes the proof. \hfill $\blacksquare$

\begin{figure}[htbp]
\begin{center}
\includegraphics[width=0.80\textwidth ]{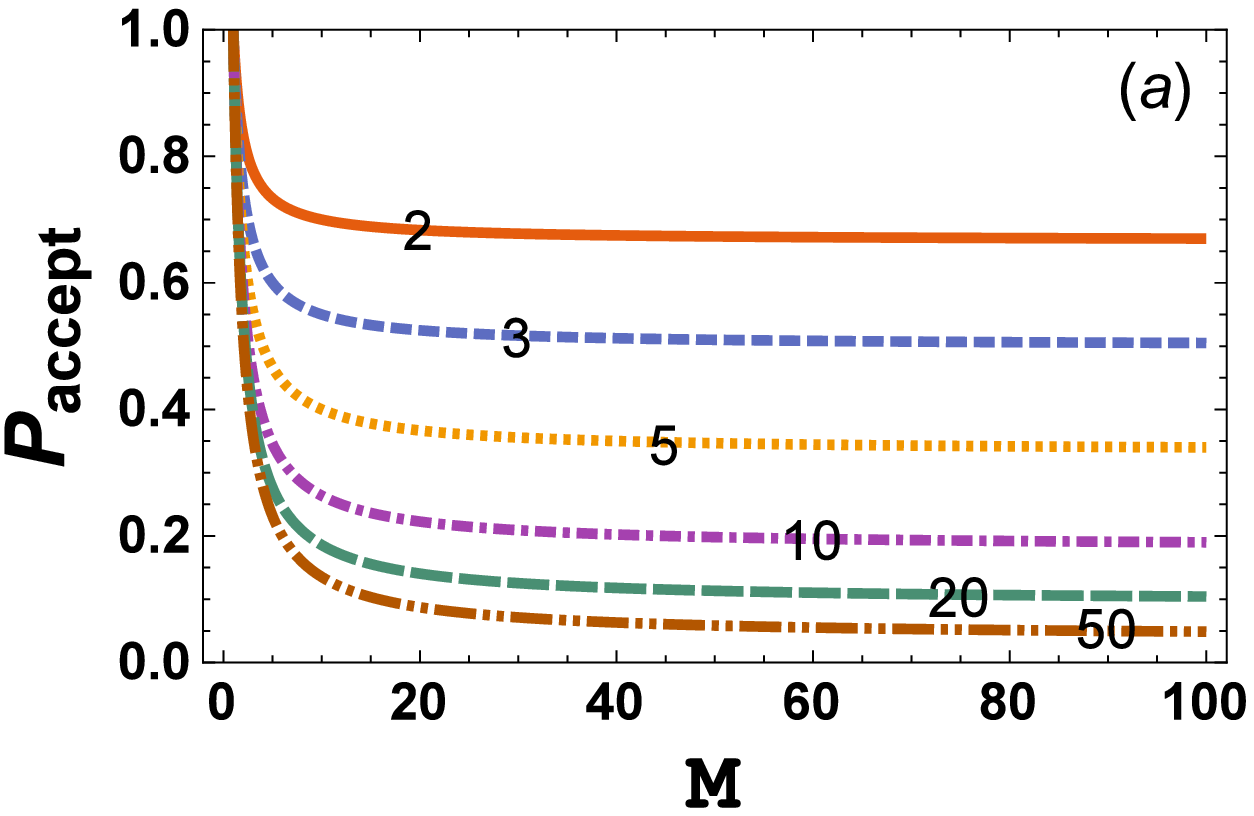}\\
\includegraphics[width=0.80\textwidth ]{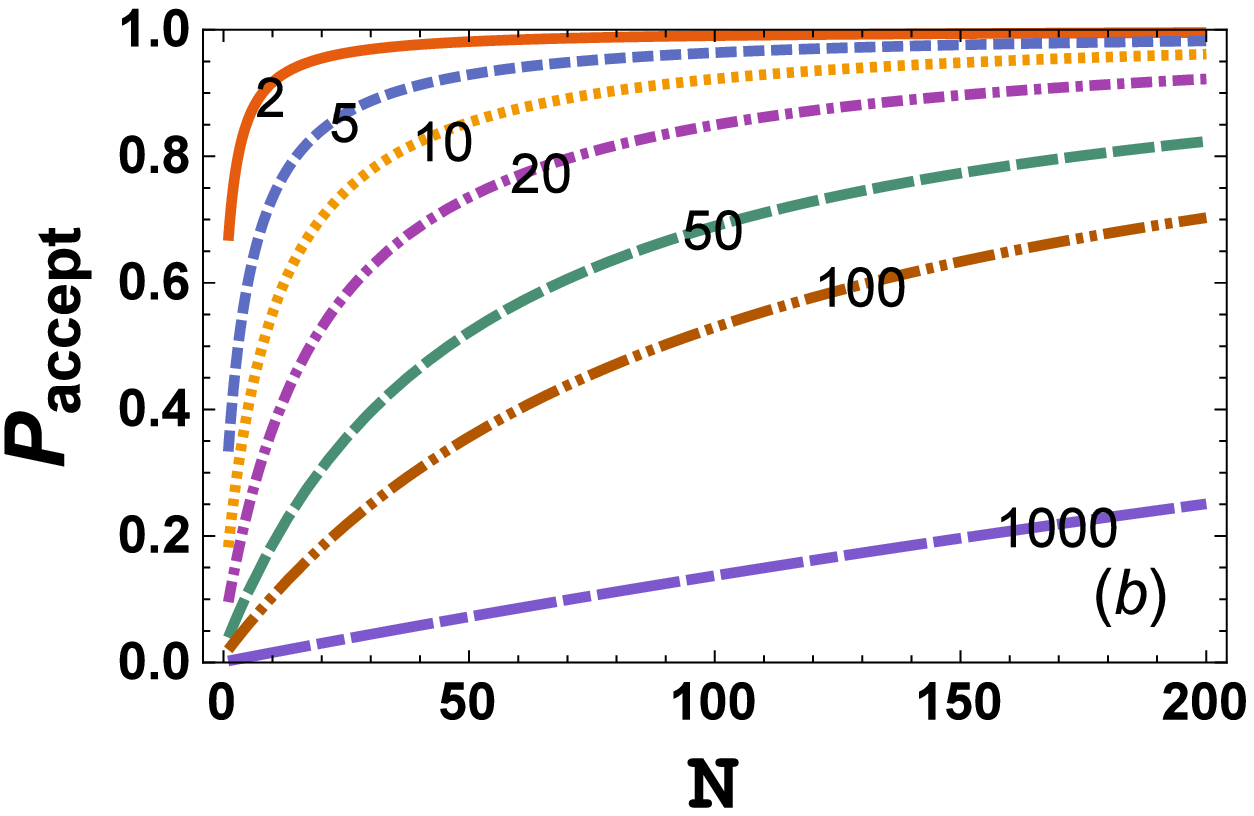}
\end{center}
\caption{(a) The false-accept probability $\mathcal{P}_{\text{accept}}$ as a function of the number of copies $M$ for different values of the dimension $d$,
when the single-photon light source is applied (that is, $N=1$); (b) $\mathcal{P}_{\text{accept}}$ as a function of the photon number $N$
of the incoming pulse for different values of $d$, when the number of copies is rather large ($M=2000$).
}\label{fig2}
\end{figure}

From Theorem \ref{T1}, we can investigate the performances of the optimal cloning attacks under various circumstances. In Fig. \ref{fig2}, we depict
two extreme but practically relevant cases: (i) when the single-photon light source is applied (that is, $N=1$), the false-accept probability
$\mathcal{P}_{\text{accept}}$ is gradually reduced with respect to the increase of the dimension $d$ for fixed $M$. In this case, $\mathcal{P}_{\text{accept}}$
becomes rather small when $M$ and $d$ are not quite large. For instance, for $M=100$ and $d=50$, $\mathcal{P}_{\text{accept}}$ is already
below $0.05$; (ii) when the number of copies $M$ is considerably large (e.g., $M=2000$), the quantum information about $|\psi\rangle$ contained in copies
is greatly diluted during the copying process \cite{Gisin1997}. On this condition, when $N>200$ and $d<100$, $\mathcal{P}_{\text{accept}}$ is always
above $0.7$, which implies that the cloning attack is rather successful. However, if $d$ is greatly increased (e.g., $d>1000$), the false-accept probability
will be significantly suppressed. This indicates that the dimension of quantum states is a critical resource for the security of authentication protocol, especially
when the $d$ is not particularly large. However, when $d$ is sufficiently large, $\mathcal{P}_{\text{UQCM}}^{\text{opt}}=\mathcal{F}_{\text{UQCM}}^{\text{opt}}$
approaches $N/M$. In this case, it is more appropriate to consider the number of copies as the resource.

\subsection{The relation between quantum cloning and challenge-estimation attacks}
In previous literature, the challenge-estimation attacks have been evaluated by B. \u{S}kori\'{c} \cite{Skoric2013a,Skoric2013b}.
In fact, the strategy of this type of attacks is as follows: the adversary (Eve) first performs a \textit{coherent} measurement $\mathcal{M}=\{M_j\}$
of all $N$ qudits. Depending on the outcome $r$, a guess $|\psi_j\rangle$ for the input state $|\psi\rangle$ is made. The average fidelity
of the estimation process is defined as
\begin{align}
\mathcal{F}_{\text{est}}(N)=\sum_j\int_{\psi}\text{tr}(M_j|\psi\rangle\langle\psi|^{\otimes N})|\langle\psi_j|\psi\rangle|^2d\psi.
\end{align}
It is worth emphasizing that the core of the problem is to find the optimal measurement and the corresponding strategy to reconstruct the guess state.
Indeed, this universal algorithm for optimal estimation of quantum states has already been established \cite{Derka1998}
and optimal fidelity has been derived \cite{Bruss1999}. From these results and the equivalence between the false-accept probability and the corresponding
fidelity, B. \u{S}kori\'{c} found the false-accept probability (per quanta) for challenge-estimation attacks is upper bounded by
\begin{align}
\mathcal{P}_{\text{est}}\leq\mathcal{P}_{\text{est}}^{\text{opt}}=\frac{N+1}{N+d},
\end{align}
where the subscript `accept' is omitted henceforth for simplicity.

In comparison with quantum cloning attacks discussed above, two mathematical observations catch our attention: (i) The upper bound of false-accept probability of
quantum cloning attacks is strictly larger than that of challenge-estimation attacks
\begin{align}
\mathcal{P}_{\text{UQCM}}^{\text{opt}}=\frac{M-N+N(M+d)}{M(N+d)}>\mathcal{P}_{\text{est}}^{\text{opt}}=\frac{N+1}{N+d},
\end{align}
(ii) When $M\rightarrow\infty$, $\mathcal{P}_{\text{UQCM}}^{\text{opt}}$ is reduced to $\mathcal{P}_{\text{est}}^{\text{opt}}$
\begin{align}
\lim_{M\rightarrow\infty}\mathcal{P}_{\text{UQCM}}^{\text{opt}}=\mathcal{P}_{\text{est}}^{\text{opt}}=\frac{N+1}{N+d}.
\end{align}

In fact, there exists a strong relation between cloning the state of a quantum system and acquiring knowledge about this state
by performing measurements \cite{Bruss1998,Bae2006}. As for our topic, these arguments in \cite{Bruss1998,Bae2006} can be
summarized by the following two theorems:

\begin{theorem}
The optimal quantum cloning attack is superior to the optimal challenge-estimation attack.
\label{T2}
\end{theorem}

\begin{figure}[htbp]
\begin{center}
\includegraphics[width=.80\textwidth]{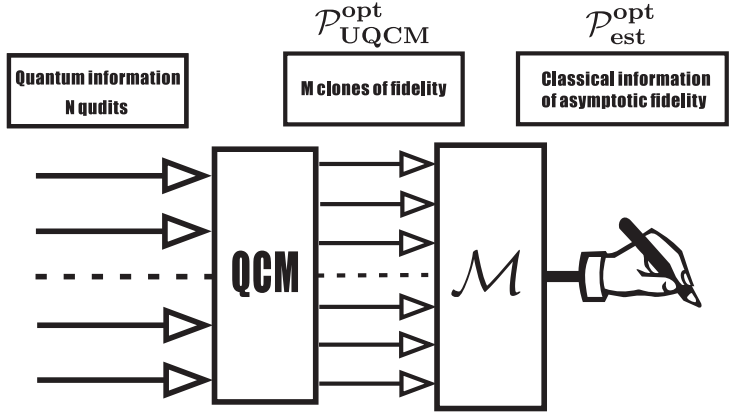} {}
\end{center}
\caption{Diagram of the information flow from quantum to classical.}
\label{fig3}
\end{figure}

\begin{theorem}
Asymptotic quantum cloning attack is equivalent to the challenge-estimation attack.
\label{T3}
\end{theorem}
The proof of Theorem \ref{T2} is relatively simple since an $M\rightarrow\infty$ cloner can be viewed as
a combination of the optimal state estimation process and a reconstruction procedure of preparing infinite
copies of the guessed state. However, this cloner is not necessary an optimal one. Mathematically,
this argument can be expressed as
\begin{align}
\mathcal{F}_{\text{QCM}}^{\text{opt}}(N,M)\geq\mathcal{F}_{\text{est}}^{\text{opt}}(N).
\end{align}
On the other hand, Theorem \ref{T3} is rather nontrivial. This argument was first proved to be true for the UQCM case \cite{Bruss1998}.
Bae and Ac\'{i}n extended this result to general cases by using the monogamy of quantum correlations and the properties of
entanglement breaking channels \cite{Bae2006}. Mathematically, this equivalence can be stated as
\begin{align}
\mathcal{F}_{\text{QCM}}^{\text{opt}}(N,M\rightarrow\infty)=\mathcal{F}_{\text{est}}^{\text{opt}}(N).
\end{align}
Hence the QCM can be regarded as a universal device transforming quantum information into
classical information \cite{Gisin1997}. The relevant information flow is illustrated in Fig. \ref{fig3}.

\subsection{Experimental considerations}
Obviously, the above theoretical findings are the guidance of experiment. However, experimental considerations are actually
more complicated. In this subsection, we make several remarks on experimental setup.

$\bullet$ \textbf{Remark 1}: To avoid being detected by the verifier via monitoring the photon number
before the verification phase, the adversary (Eve) can trace out $M-N$ copies of the output state $\mathcal{L}\left(\psi^{\otimes N}\right)$ and
resend the remaining $N$ copies back to Alice.

$\bullet$ \textbf{Remark 2}: In fact, the upper bound of the false-accept probability for a single copy $\mathcal{P}_{\text{UQCM}}^{\text{opt}}$
is a function of the triple $\{N,M,d\}$. A closer look at this function reveals that $\mathcal{P}_{\text{UQCM}}^{\text{opt}}$
is a monotone increasing function of $N$, but monotone decreasing with respect to $M$ and $d$. Therefore,
Alice and Eve should take different operations according to their targets (see Table \ref{Table1}).

\begin{table}[!hbp]
\centering
\begin{tabular}{c c c c c}
\hline
\hline
Identity & Target & $N$ & $M$ & $d$ \\
\hline
Alice & $\mathcal{P}\searrow$ & $N\searrow$ & No control & $d\nearrow$ \\
Eve & $\mathcal{P}\nearrow$ & No control & $M\searrow$ & No control \\
\hline
\hline
\end{tabular}
\caption{Targets of Alice and Eve and their corresponding manipulations.}
\label{Table1}
\end{table}

In the protocol described in Sect. \ref{sec2.2}, a specific challenge state $|\psi\rangle$ is usually applied $m$ times. Thus, the
expected total number of photons is $Nm\mathcal{P}_{\text{UQCM}}$. Due to the photon loss and detection imperfections, actually
if $\mathcal{P}_{\text{UQCM}}\geq1-\varepsilon$ the attacker will be accepted by the authentication protocol. Since the total detection
precess can be viewed as a sequence of $Nm$ independent yes/no experiments, the total false-accept probability obeys the
binomial distribution

\begin{align}
\mathcal{P}_{\text{total}}&=\sum_{k=\lceil Nm(1-\varepsilon) \rceil}^{Nm}
\left(\begin{matrix}
Nm \\
k \\
\end{matrix}\right)
\mathcal{P}_{\text{UQCM}}^k(1-\mathcal{P}_{\text{UQCM}})^{Nm-k}\nonumber\\
&=I_{\mathcal{P}}(n+1,Nm-n)
\end{align}
where $n=\lfloor Nm(1-\varepsilon) \rfloor$ and $I_x(a,b)$ is the so-called regularized incomplete beta function.
We can prove that $\mathcal{P}_{\text{total}}$ is a monotone increasing function of $\mathcal{P}_{\text{UQCM}}$,
and thus the results in Table \ref{Table1} can also be adopted to improve the security of the authentication protocol
(see Appendix \ref{appendix2} for more details).

$\bullet$ \textbf{Remark 3}: It is important to note that some imperfect single photon sources are unitized
in realistic experiments \cite{Goorden2014}. On this occasion, the number of photons in a single pulse
is not fixed and follows a certain probability distribution (e.g., Poisson distribution for weak coherent states).
Accordingly, the verification threshold should be modified to $N_{\text{av}}m(1-\varepsilon)$, where $N_{\text{av}}=\mathbb{E}_{N}[N]$ denotes
the average photon number. Meanwhile, we should average the false-accept probability over such a distribution.
In this case, we have
\begin{align}
\mathcal{P}_{\text{av}}&\leq\mathbb{E}_{N}\frac{M-N+N(M+d)}{M(N+d)}\nonumber\\
&=\mathbb{E}_{N}\left[1+\frac{d-1}{M}-\frac{(d-1)(M+d)}{M(N+d)}\right]\nonumber\\
&=1+\frac{d-1}{M}-\mathbb{E}_{N}\left[\frac{(d-1)(M+d)}{M(N+d)}\right]\nonumber\\
&\leq1+\frac{d-1}{M}-\frac{(d-1)(M+d)}{\mathbb{E}_{N}[M(N+d)]}\nonumber\\
&=\frac{M-N_{\text{av}}+N_{\text{av}}(M+d)}{M(N_{\text{av}}+d)},
\end{align}
where Jensen's inequality and the convexity of $1/x$ are used. Note that the only difference is the substitution $N\Rightarrow N_{\text{av}}$.

Actually, when the detection efficiency $\eta$ is close to 1, the ideal single-photon source is indeed a better
choice for the verifier since the false-accept probability $\mathcal{P}_{\text{accept}}$ is a monotone increasing
function of $N_{\text{av}}$. However, in the language of the experiment [10], the decision of acceptance or rejection is totally
determined by the difference of probability distribution of the number of photodetections. Therefore, if the detection
efficiency is rather low, the choice of $N_{\text{av}}=1$ will have a consequence that we can hardly distinguish between the true and
random PUF since in this case the number of photodetections is too low and thus it is difficult to discriminate
the corresponding probability distribution. This situation will result in a lower probability of
detecting an adversary who tries to pretend to have access to the PUF. Hence the verifier can appropriately adjust
the average photon number to a moderate value (e.g., $N_{\text{av}}=230\pm40$ [10]), but note that on this condition
the false-accept probability $\mathcal{P}_{\text{accept}}$ is greatly increased.
Therefore, a trade-off relation between the average photon number and the detection
efficiency should be taken into considerations in a realistic experiment.

\section{Comparison with QKD protocols and other cloning attacks}\label{sec4}
For QKD protocols, the relationship between the no-cloning theorem and the security of quantum
cryptography was already clarified in the first protocol, that is, the BB84 protocol \cite{Bennett1984}.
In fact, various types of quantum cloning machines have been designed to analyze the security of QKD protocols.
Particularly, quantum cloning attacks are proved to be the optimal incoherent attacks for BB84 protocol,
six-state protocol and continuous-variable protocols \cite{Scarani2005,Fan2014}. Although there is no
equivalence relation between the optimal cloning and optimal eavesdropping, their relationship have been shown
to be strong and fruitful in the framework of quantum cryptography.

It is worth noting that the coding of BB84 and six-state protocols has been generalized to
larger dimensional quantum systems \cite{Bechmann2000a,Bechmann2000b,Cerf2002}.
Specifically, the spatial degrees of freedom is exploited in high-dimensional QKD experiments \cite{Walborn2006,Etcheverry2013}.
Similarly, the high-spatial-dimension states of light are also employed in the PUF-based quantum authentication experiment \cite{Goorden2014},
where the number of controlled modes $d=1100\pm200$. Remarkably, the SLM has played a critical role in creating
the desirable challenge states. Intriguingly, in the context of quantum cryptography, the SLM is also used to
produce the $d$-dimensional states, which are of the form \cite{Etcheverry2013}
\begin{align}
|\Psi\rangle=\frac{1}{d}\sum_{-l_d}^{l_d}e^{i\phi_l}|l\rangle ,
\label{PQCM}
\end{align}
where $l_d=(d-1)/2$, $|l\rangle$ form the logical basis in the $d$-dimensional Hilbert space of the transmitted photons,
and $\phi_l$ are the phases introduced by the SLM \cite{Neves2005}.

Obviously, one possible scenario is that the adversary may acquire \textit{partial} information of the challenge states
through the configuration of the SLM. For instance, the adversary may know that a specific challenge state is of the form (\ref{PQCM}).
In this case, the phase-covariant quantum cloning machine (PQCM) for qudits can be employed to launch an attack against
the quantum authentication systems, which is more powerful than the UQCM attacks \cite{Fan2003}. For example,
the optimal fidelity of $1\rightarrow 2$ PQCM is larger than that of UQCM \cite{Fan2003}
\begin{align}
\mathcal{F}_{\text{PQCM}}^{\text{opt}}(1,2)&=\frac{1}{d}+\frac{1}{4d}\left(d-2+\sqrt{d^2+4d-4}\right)\nonumber\\
&>\mathcal{F}_{\text{UQCM}}^{\text{opt}}(1,2)=\frac{d+3}{2(d+1)}.
\end{align}
Therefore, the leakage of the configuration information of the SLM will greatly compromise the security of
the PUF-based quantum authentication systems. At the same time, other quantum cloning machines such as
PQCM can be exploited to attack the the authentication systems.

\section{Conclusions}\label{sec5}
In this work, we systematically studied the quantum cloning attacks on the PUF-based quantum authentication systems.
First, we obtained the analytical formula of the false-accept probability by use of the universal quantum cloning machines
and proved that optimal quantum cloning attack outperforms the so-called challenge-estimation attacks. Remarkably, we have
established the relationship between these two types of attacks and the information flow in the whole process is clarified.
Moreover, from the experimental perspective, a trade-off between the average photon number and the detection efficiency
is discussed in detail. Finally, an explicit comparison is made between QKD protocols and quantum authentication protocols
and other possible cloning attacks are illustrated.

In view of these findings, a set of topics can be pursued as the future research directions:
(i) A detailed and integrated security analysis of the PUF-based quantum authentication protocols
is still missing. Other sorts of attacks may prove to be more efficient and the corresponding
defending methods will be proposed. Such a positive interaction between attacking and defending
will enable us to gain deeper insight into this subject. Interesting inspirations may come from
the field of quantum cryptography; (ii) Since QKD protocols can be viewed as some form of
``quantum PUFs'' \cite{Plaga2012}, the investigation of the security of quantum cryptography
may be unified into the framework of PUFs. This new direction seems promising and the development
of secure ``quantum PUFs'' capable of tolerating realistic imperfections is of both practical
and fundamental importance.

\begin{acknowledgements}
We are grateful to the Foundation of President of the China Academy of Engineering Physics under Grant
No. 2014-1-100.
\end{acknowledgements}

\appendix
\section{Attack models}\label{appendix1}
\begin{figure}[htbp]
\begin{center}
\includegraphics[width=0.80\textwidth ]{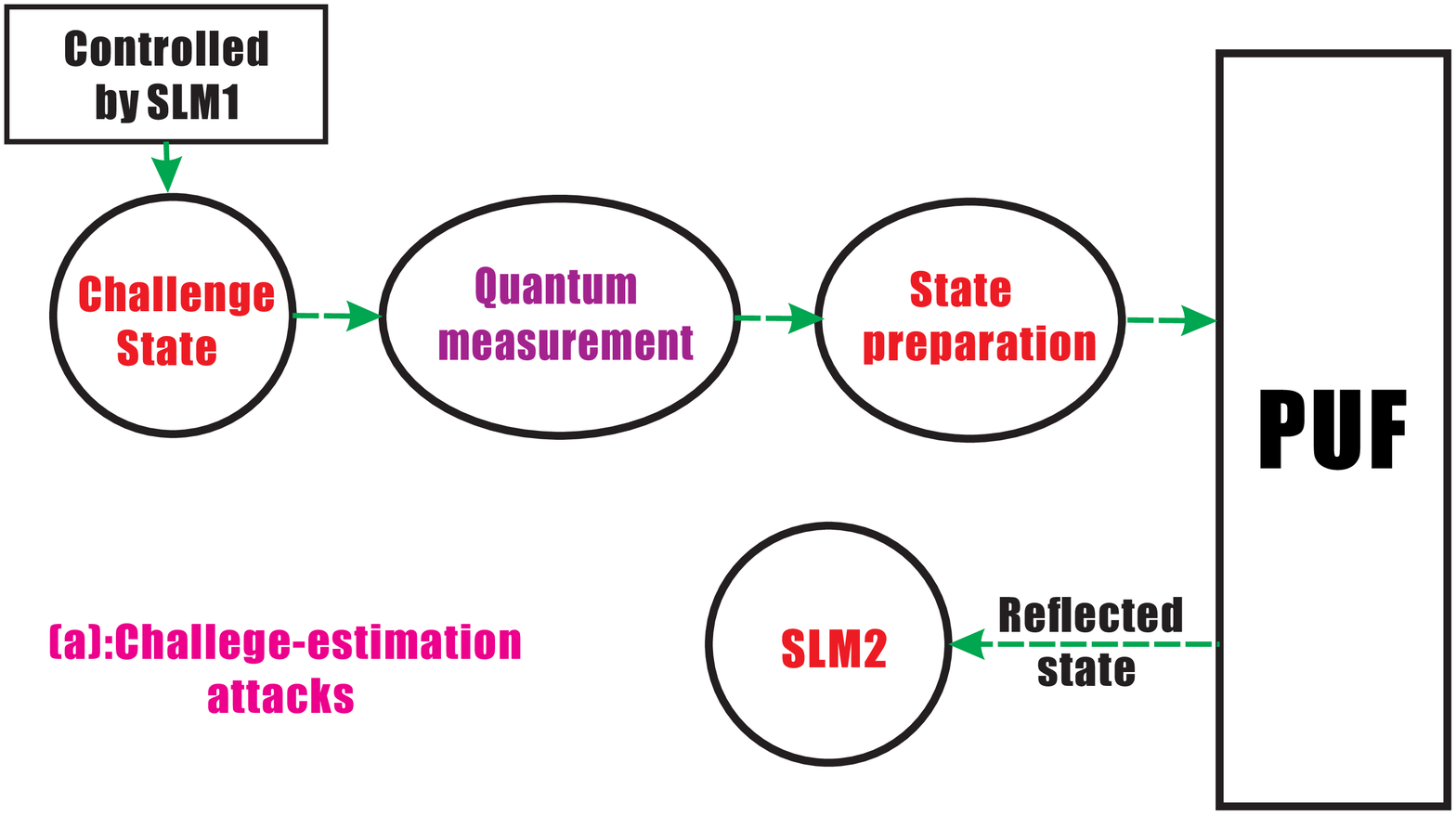}\\
\includegraphics[width=0.80\textwidth ]{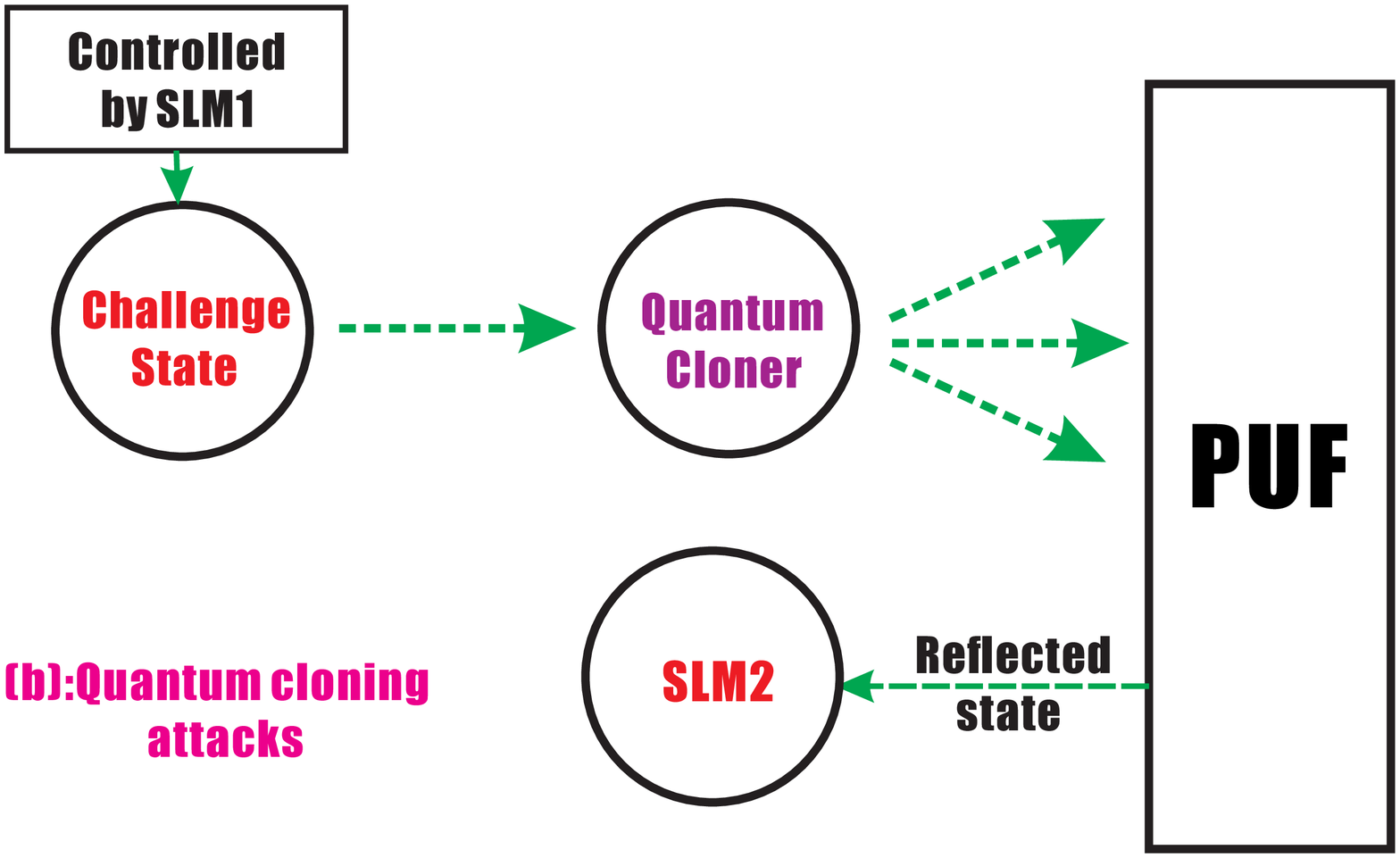}
\end{center}
\caption{ The comparison of the attack models of challenge-estimation and quantum-cloning.
}\label{fig4}
\end{figure}

First, we would like to emphasize that not only in our attack model but also \cite{Skoric2012,Skoric2013a,Skoric2013b} the target of
the attacks is only \textbf{the challenge state}. In particular, in the context of the experiment \cite{Goorden2014},
the false PUF key is actually imitated by sending random challenges to the true key (see caption of Fig. 3 in \cite{Goorden2014}).
In fact, in the language of \cite{Goorden2014}, the verification phase goes as follows: (i) the challenge state is determined by the configuration of SLM1;
(ii) the PUF \textbf{automatically} performs the unitary transformation R; (iii) the SLM2 functions as the projection measurement
$(R|\psi\rangle)^\dagger=\langle\psi|R^\dagger$. As shown in Fig. \ref{fig4}, it is evident that the difference of the two attack models is that:

\textbf{Challenge-estimation attacks}: Quantum measurement strategy + State preparation

\textbf{Quantum cloning attacks}: Implantations of quantum cloning machines

Since the attack model described in \cite{Skoric2012,Skoric2013a,Skoric2013b} adopts the so-called ``measurement and prepare'' strategy, this attack model
is classified to the (strongest) type of classical attacks. However, in our attack model, no measurement is involved except
for the projection of SLM2. More precisely, we preform the quantum attack against the challenge state by employing quantum
cloning machines. Actually, the justification of our statement and \cite{Goorden2014} relies on the following \textbf{equivalence relationship}:
\begin{center}
\textbf{Correct challenge state + Random PUF key}\\
$\Updownarrow$\\
\textbf{Random challenge state + True PUF key}
\end{center}
since an unclonable physical key (PUF) can be viewed as a unitary transformation. Note that the \textbf{only key difference}
lies in that the ``measure and prepare'' strategy is \textbf{replaced} by ``quantum cloning'' scenarios in our model.

\section{The cumulative binomial probability function}\label{appendix2}
In general, if the random variable $X$ follows the binomial distribution with $N$ trials and success probability $p$ for each trial,
the cumulative distribution function can be expressed as
\begin{align}
F(n;N,p)&=P({X\leq n})=\sum_{k=0}^{n}
\left(\begin{matrix}
N \\
k \\
\end{matrix}\right)p^k(1-p)^{N-k}\nonumber\\
&=I_{1-p}(N-n,n+1)\nonumber\\
&=(N-n)
\left(\begin{matrix}
N \\
n \\
\end{matrix}\right)\int_0^{1-p}t^{N-n-1}(1-t)^ndt
\end{align}
where $I_x(a,b)$ is the the regularized incomplete beta function
\begin{align}
I_{x}(a,b)=\frac{B(x;a,b)}{B(a,b)},
\end{align}
$B(a,b)$ is the beta function, and $B(x;a,b)$ is the incomplete beta function.
In our context, the probability of obtaining more successes than $n$ observed in a binomial distribution is
\begin{align}
P({X> n})=1-F(n;N,p)
=I_{p}(n+1,N-n),
\end{align}
where we have used the relation $I_x(a,b)+I_{1-x}(b,a)=1$. Note that the partial derivative of $B(x;a,b)$ with respect to $x$ is
\begin{align}
\frac{\partial B(x;a,b)}{\partial x}=(1-x)^{b-1}x^{a-1}.
\end{align}
Therefore, we have
\begin{align}
\frac{\partial I_{p}(n+1,N-n)}{\partial p}=\frac{(1-p)^{N-n-1}p^n}{B(N-n,n+1)}>0,
\end{align}
which implies that $I_{p}(n+1,N-n)$ is a monotone increasing function of $p$.



\end{document}